\documentclass[fleqn,usenatbib]{mnras}

\usepackage{newtxtext,newtxmath}

\usepackage[T1]{fontenc}

\DeclareRobustCommand{\VAN}[3]{#2}
\let\VANthebibliography\thebibliography
\def\thebibliography{\DeclareRobustCommand{\VAN}[3]{##3}\VANthebibliography}

\usepackage{graphicx}	
\usepackage{amsmath}	
\usepackage{color}
\usepackage{comment}
\usepackage{lscape} 
\usepackage{longtable}
\usepackage{supertabular}
\usepackage{booktabs}
\usepackage{scrextend}

\newcommand{\astfootnote}[1]{%
\let\oldthefootnote=\thefootnote%
\setcounter{footnote}{0}%
\renewcommand{\thefootnote}{\textcolor{black}{\fnsymbol{footnote}}}%
\footnote{#1}%
\let\thefootnote=\oldthefootnote%
}

\definecolor{amethyst}{rgb}{0.6, 0.4, 0.8}
\definecolor{orange}{rgb}{0.89, 0.26, 0.2}

\title[Machine Learning to identify ICL and BCG]{Machine Learning to identify ICL and BCG in simulated galaxy clusters}

\author[I. Marini et al.]{
I. Marini,$^{1,2,3,4}$\thanks{E-mail: ilaria.marini@inaf.it} S. Borgani,$^{1,2,3,4}$ A. Saro,$^{1,2,3,4}$ G. Murante,$^{2,3}$ G. L. Granato,$^{2,5,3}$   \\~\\
{\rm {\LARGE  C. Ragone-Figueroa,$^{5,2}$ G. Taffoni,$^{2}$ }}
\\
    $^1$ Astronomy Unit, Department of Physics, University of Trieste, via Tiepolo 11, I-34131 Trieste, Italy\\
    $^2$ INAF -- Osservatorio Astronomico di Trieste, via G. B. Tiepolo 11, I-34143, Trieste, Italy\\
    $^3$ IFPU -- Institute for Fundamental Physics of the Universe, Via Beirut 2, 34014 Trieste, Italy\\  
    $^4$ INFN -- Sezione di Trieste, Trieste,  Italy\\
    $^5$ Instituto de Astronom\'ia Te\'orica y Experimental (IATE), Consejo Nacional de Investigaciones Cient\'ificas y T\'ecnicas de la\\
Rep\'ublica Argentina (CONICET), Universidad Nacional de C\'ordoba, Laprida 854, X5000BGR, C\'ordoba, Argentina\\
}

\date{Accepted 2022 May 30. Received 2022 April 21; in original form 2022 March 04}

\pubyear{2022}

\def \msun{{$M_{\odot}$}}
\def \m200{{$M_{200}$}}
\def \r200{{$R_{200}$}}
\def \v200{{$V_{200}$}}
\def \dispbcg{{$\sigma_\mathrm{BCG}$}}
\def \dispicl{{$\sigma_\mathrm{ICL}$}}

\usepackage{makecell}
\usepackage{graphicx}

\begin{document}
\label{firstpage}
\pagerange{\pageref{firstpage}--\pageref{lastpage}}
\maketitle

\begin{abstract}
Nowadays, Machine Learning techniques offer fast and efficient solutions for classification problems that would require intensive computational resources via traditional methods. We examine the use of a supervised Random Forest to classify stars in simulated galaxy clusters after subtracting the member galaxies. These dynamically different components are interpreted as the individual properties of the stars in the Brightest Cluster Galaxy (BCG) and IntraCluster Light (ICL). We employ matched stellar catalogues (built from the different dynamical properties of BCG and ICL) of 29 simulated clusters from the DIANOGA set to train and test the classifier. The input features are cluster mass, normalized particle cluster-centric distance, and rest-frame velocity. The model is found to correctly identify most of the stars, while the larger errors are exhibited at the BCG outskirt, where the differences between the physical properties of the two components are less obvious. We investigate the robustness of the classifier to numerical resolution, redshift dependence (up to $z=1$), and included astrophysical models. We claim that our classifier provides consistent results in simulations for $z<1$, at different resolution levels and with significantly different subgrid models. The phase-space structure is examined to assess whether the general properties of the stellar components are recovered: (i) the transition radius between BCG-dominated and ICL-dominated region is identified at $0.04$ \r200; (ii) the BCG outskirt ($> 0.1$ \r200) is significantly affected by uncertainties in the classification process. In conclusion, this work suggests the importance of employing Machine Learning to speed up a computationally expensive classification in simulations. 
\end{abstract}

\begin{keywords}
methods: data analysis -- methods: statistical -- galaxies: stellar content
\end{keywords}



\section{Introduction}
In recent years, the diffuse stellar envelope observed in groups and clusters of galaxies, called IntraCluster Light (ICL), has assumed a prominent place in the study of structure formation. This visible tracer exhibits properties that are rather peculiar, distinct from the other stars confined in their constituent member galaxies \citep[e.g.][and references therein]{contini2021origin,montes2022faint}. Both theoretical and observational evidences \citep[][just to quote a few]{murante2004diffuse,murante2007importance, puchwein2010intracluster, mihos2016burrell, montes2018intracluster, montes2019intracluster, spavone2020fornax,kluge2020structure} have been gathered on the origin and evolution of this component. Recent findings have suggested that the ICL distribution follows the global potential well of the host galaxy cluster \citep[e.g.,][]{montes2019intracluster,alonso2020intracluster,canas2020stellar} and thus, it can be used as a luminous tracer for dark matter, highlighting its importance in the context of structure formation. 

Observationally constraining the properties of the ICL is troublesome, as it requires both deep and wide observations of spatially extended low-surface brightness regions in the sky, other than a top-level data processing pipeline to avoid spurious contamination from other sources. To further complicate the scenario, the evolution of the ICL is tightly connected to the build-up of the Brightest Cluster Galaxy (BCG), i.e., the central galaxy in a cluster, which sits at the centre of the cluster gravitational potential. Both the spatial extent and luminosity curves of the two components smoothly merge, leaving no trace of the transition \citep{bender2015structure,kluge2020structure}. Therefore, the separation of the ICL from the BCG is performed in several (often laborious) ways. Some studies \citep[e.g.][]{kluge2020structure,spavone2020fornax} identify the ICL as the excess of light with respect to a de Vaucouleurs profile or a double Sérsic decomposition, while often it is preferred to perform a simple cut in surface brightness \citep[][]{mihos2016burrell,montes2018intracluster}.  In this regard, several studies \citep[e.g.][and references therein]{contini2022transition,montes2022faint} have discussed the role of the transition radius, i.e. the cluster-centric distance at which the ICL component starts dominating the stellar component. Due to the variety of methods employed to estimate the ICL contribution, the value of this transition radius may depend on the adopted method of ICL identification. From the observational side,  typical values of the  transition radius are around $60-80$ kpc \citep[][]{montes2021buildup,gonzalez2021discovery}, thus in line with results from earlier works \citep[][]{zibetti2005intergalactic,gonzalez2007census,seigar2007intracluster,iodice2016fornax}. These values slightly increase for other analyses, such as those presented by \citet{zhang2019dark}, who concluded that the transition from the BCG to the ICL is just outside $100$ kpc, or by \citet{chen2021sphere} who found values ranging in the interval $70-200$ kpc. Results based on simulations \citep[e.g.,][]{contini2021brightest,contini2021origin,contini2022transition} agree with these observational results, and indicate that the transition radius is independent of both BCG+ICL and halo masses, with typical values of $60\pm40$ kpc, if similarly derived from profile fitting. Usually, this technique requires the assumption of a double/triple S\'{e}rsic profile \citep[][]{sersic1963influence} or a composition of different profiles such as the Jaffe profile \citep[][describing the BCG distribution]{jaffe1983simple} and NFW profile \citep[][]{navarro1997universal} for the ICL.

To our advantage, in simulations, one can exploit the full 6D phase-space information available on star particles to investigate the properties of the ICL and BCG \citep[][]{dolag2010dynamical,remus2017outer}. It is in this direction that \citet{dolag2010dynamical} have invested their effort in designing a classification algorithm applicable to the star particles in the main halo of simulated clusters and groups according to their properties in phase-space. The assumptions underlying this method derive from the study of the velocity distribution of star particles which exhibit a bimodal distribution that can be associated with two distinct dynamical components. Combining this information with an unbinding procedure leads to separation into a central BCG (more compact and dynamically cold) and a diffuse ICL. Although this method should not be regarded as a procedure with outputs immediately comparable to observations, it provides us with the dynamical information associated with each component. In other words, we expect this technique to convey information on the physical properties of both stellar components, to complement the observational data. 

To our disadvantage, the large volume of data to classify in state-of-the-art simulations requires intensive computational effort. To overcome this limitation, the analysis presented in this paper aims at reproducing a similar classification method adopting Machine Learning (ML) techniques that often prove to be less computationally expensive and more efficient than traditional methods. An automated methodology for efficiently classifying the stellar components can be an essential ingredient to facilitate the use of these tools in nowadays analyses. To this end, we build a Random Forest classifier to recognize the label of a star particle solely basing the decision on the specific features of each particle. This method is widely employed in ML problems for its versatility and its performance with high-dimensional data. One essential benefit is that the computational cost of Random Forest models does not depend significantly on the size of the training set, given that it scales logarithmically. Additionally, the predictions are straightforward to interpret, while it is also extremely easy to measure the relative importance of each feature in the predictions.

The paper is structured as follows. In Section \ref{sec:simulations} we present the synthetic cluster set on which we train, cross-validate, and test the classifier. Furthermore, we include a description of the traditional method used to calibrate the ML model. Section \ref{sec:RF} describes the model and its caveats; in Section \ref{sec:results}, we discuss the achieved classification performance with distinct clusters and assess the reliability of the model to recover the true label, as identified by the ICL-Subfind. Finally, we present our conclusions in Section \ref{sec:conclusions}.

\section{Simulations}
\label{sec:simulations}
The ML algorithm is trained, cross-validated, and tested on 29 clusters from a set of cosmological hydrodynamical simulations called DIANOGA. These simulations were carried out with the GADGET-3 code, a modified version of the GADGET-2 tree-PM smoothed particle hydrodynamics (SPH) public code \citep{springel2005cosmological}. The major changes include a higher-order kernel function, a time-dependent artificial viscosity model, and a time-dependent artificial conduction scheme. 

The 29 simulated clusters (for simplicity called D1, D2,..., D29) are the result of zoom-in simulations centered on the most massive galaxy clusters evolved in a lower-resolution N-body parent box of 1 h$^{-3}$ Gpc$^{3}$ volume with the inclusion of baryons. The cosmological model is a $\Lambda$ Cold Dark Matter (CDM) with the following parameters $\Omega_M=0.24$, $\Omega_b =0.04$, $n_s = 0.96$, $\sigma_8=0.8$ and $H_0 = 72$ km s$^{-1}$ Mpc$^{-1}$. These clusters represent the 24 most massive clusters in the parent box with masses $M_{200} \in [0.8 - 2.7] \times 10^{14}$ h$^{-1}$ M$_{\odot}$ and 5 isolated groups with M$_{200}$ within [1-4] $\times 10^{14}$ h$^{-1}$ M$_{\odot}$. In the high-resolution regions, the DM particle mass is $m_\textrm{DM} = 8.3 \times 10^8$ h$^{-1}$ M$_{\odot}$ and the initial mass of the gas particle is $m_\textrm{gas} = 3.3 \times 10^8$ h$^{-1}$ M$_{\odot}$. The Plummer equivalent gravitational softening for DM particles is set to $\epsilon =$ 5.75 h$^{-1}$ kpc. The gravitational softening lengths of gas, star, and black hole particles are 5.75 h$^{-1}$ kpc, 3 h$^{-1}$ kpc and 3 h$^{-1}$ kpc, respectively. Several subgrid models included in the simulations treat the unresolved baryonic physics of the simulations. Details can be found in \cite{ragone2018bcg} and references therein.

\subsection{Halo finder}
    Self-bound structures (i.e. our "bonafide" galaxies) are identified by running Subfind \citep{springel2001populating,dolag2009substructures} in the catalogue of group particles compiled by Friends of Friends (FoF) with link length b = 0.16 in units of the mean interparticle distance. The algorithm selects out local density maxima, identified in a geometrical way. Particles not gravitationally bound to such maxima are discarded. A substructure is considered resolved if it contains at least 50 DM and/or star particles. 
    
    In this work, rather than considering these newly found galaxies and their properties, we employ the catalogue to discard all the particles associated with the satellite substructures. We keep only the star particles in the main halo (defined as the central subhalo of a given group), which contains, besides the BCG, all particles that were not associated with any other subhalo.

\subsection{The ICL separation by Subfind}
\begin{figure*}
    \includegraphics[scale=0.5,angle=0.0]{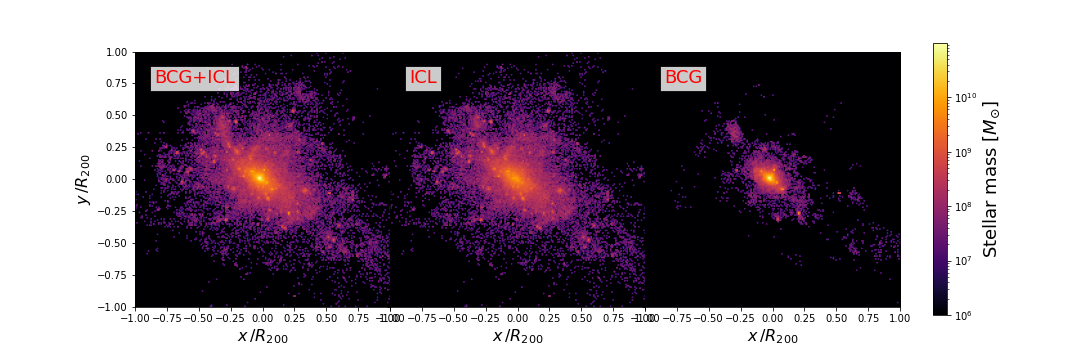}
    \caption{Mass-weighted maps of the stellar components as identified by the ICL-Subfind for one of the clusters in our simulations. \emph{Left: } BCG and ICL hosted in the main halo of the cluster. \emph{Center: } Stars associated with the ICL component. \emph{Right: } Stars bound to the BCG. The stellar component is dominated by the ICL diffuse component.}
    \label{fig:only_true_maps}
\end{figure*}

\begin{figure*}
    \includegraphics[scale=0.45,angle=0.0]{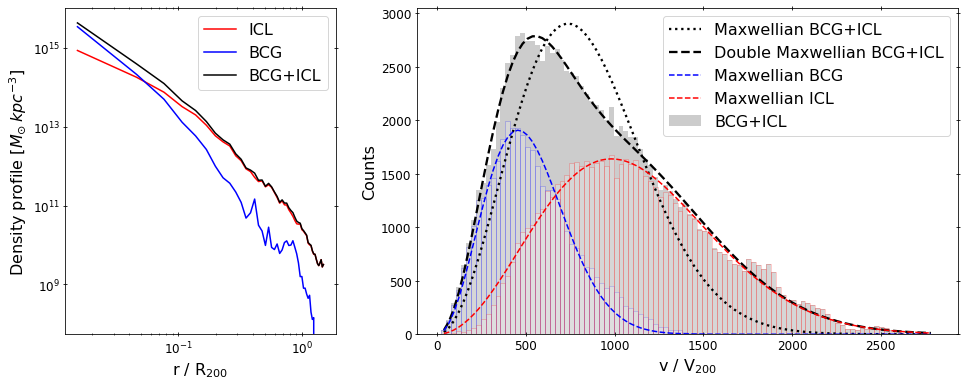}
    \caption{Observed properties in the stellar populations identified by the ICL-Subfind for the same cluster in Fig. \ref{fig:only_true_maps}. \emph{Left: } Main halo (black), ICL (red), and BCG (blue) stellar density profiles.  \emph{Right:} Velocity histograms of the main halo (black), ICL (red), and BCG (blue). We also report the single (dashed black) and double Maxwellian (dotted black) best-fit curves for the entire data set. Red and blue dashed lines are showing the individual Maxwellian distributions which are associated with the unbound ICL ($\sigma_\textrm{ICL}^{\star} = 1002$  km s$^{-1}$) and bound BCG ($\sigma_\textrm{BCG}^{\star} = 484$  km s$^{-1}$).}
    \label{fig:onlytrue}
\end{figure*}

To determine the dynamical distinction between ICL and BCG in the stellar envelope of the main halo of each group, we implement a modified version of the halo finder Subfind, which for clarity we will call "ICL-Subfind". This division is performed once all star particles of the central group are isolated, by firstly subtracting the stars associated with the member galaxies of the group with the standard version of Subfind (presented in the previous section). The details of this technique are presented in \cite{dolag2010dynamical}, here we will only provide a brief description.

The algorithm identifies the single star particles in the main halo of clusters as either bound to the BCG or ICL, solely applying a dynamical criterion. The underlying assumption is that the two velocity distributions of stars belonging to the ICL and the BCG can be each fitted by a Maxwellian shape so that the overall velocity distribution of stars is described by a double Maxwellian of the following form:
\begin{equation}
\label{eq:double_maxw}
    N(v) = k_1 v^2\exp\left(-\frac{v^2}{2\sigma_1^2}\right)+k_2 v^2\exp\left(-\frac{v^2}{2\sigma_2^2}\right) .
\end{equation} The diffuse ICL is associated with the Maxwellian yielding the largest velocity dispersion, in contrast, the BCG, having colder dynamics, populates the distribution at lower dispersion. 

To assign each star particle to either one of the two dynamical components, the algorithm follows an unbinding procedure, by iteratively computing the gravitational potential given by all particles within a sphere whose radius is initially equal to a fraction of the virial radius. In this framework, we compare each particle's kinetic energy with the potential energy (at the particle position) given by this spherical mass distribution. If the particle's kinetic energy is higher, then the particle is defined as "unbound" (and "bound" otherwise). Performing this operation on all star particles identifies two stellar populations which are then separately fitted with a single Maxwellian. If the best-fit parameters of the double Maxwellian match those obtained from the single initial Maxwellians, then the procedure is completed, otherwise, the radius of the sphere is changed and the computation is remade. Notice that the radius is usually adjusted to match the value of the BCG components, but provided that the algorithm does not converge, then a second attempt is made with the ICL component. The radius is varied so as to decrease (increase) the spherical mass distribution according to whether the BCG velocity dispersion is too high (low) compared to the expected result from the initial fit. The iterative procedure stops when the ratio of the expected velocity dispersion over the fitted one differs from less than a given threshold value and thus, one obtains the label "ICL" or "BCG" for each star particle in the main halo. Nevertheless, the algorithm may not necessarily converge if the number of iterations exceeds a threshold value provided by the user. We will see that this is the case for some of our systems.

As a proof of concept, Fig. \ref{fig:only_true_maps} illustrates the outcome of the labelling in the total stellar population of the main halo (left panel) from the stellar map weighted on the particle masses of one of the clusters in our simulation. By applying the ICL-Subfind algorithm, we can separately study the characteristics of the ICL (central panel) and BCG population (right panel). Similarly, Fig. \ref{fig:onlytrue} shows the resulting density profiles (left panel) and velocity histograms (right panel) of the same cluster. Radius and velocity are normalized by \r200 and \v200\footnote{We can define $R_{\Delta}$ as the radius encompassing a mean overdensity equal to $\Delta$ times the critical density of the universe $\rho_c$. Equivalently, we define the circular velocity of the virion $V_{200} = \sqrt{G M_{200}/R_{200}}$}. The colour-coded legend is common to both panels: stars in the main halo in black, ICL in red, and BCG in blue. The density profiles show that the BCG mostly resides in the central regions, while the ICL extends to larger distances, in fact dominating the stellar component in the outskirts. In the right panel, we plot the histograms of the main halo star particles velocity distribution (in black) and the single BCG (blue) and ICL (red). We observe that a single Maxwellian (best fit $\sigma^{\star} = 867$ km s$^{-1}$), represented by the dotted black line, does not provide a good fit to the particle distribution. On the other hand, when the fitting procedure is attempted with a double Maxwellian (black dashed line), the agreement is much more evident. The single Maxwellian associated with both the BCG and ICL are also reported with the dashed lines. The diffuse ICL is associated with the Maxwellian with the larger velocity dispersion ($\sigma^{\star}_\textrm{ICL} = 1002$ km s$^{-1}$), in contrast, the BCG, having colder dynamics, populates the distribution at lower dispersion ($\sigma_\textrm{BCG}^{\star} = 484$ km s$^{-1}$). Furthermore, in \citet{dolag2010dynamical} it was tested that a triple Maxwellian does not improve the results in most cases.

\section{Random Forest}
\label{sec:RF}
Given the nature of the ICL-Subfind algorithm, based on well-defined properties of the stellar components, its action may be also replicated by a ML model in a faster and more efficient way. Our goal is to provide an alternative classification method for identifying stars in the main halo according to several features that are crucial in the use of the former method. To achieve this, we design a supervised classification method, based on the Random Forest classifier \citep[based on][]{scikit-learn}, to which we feed a feature vector representative of the classes we are predicting (i.e., BCG and ICL). Examples of input features we tested are the potential and total energy of each particle, the particle age, mass, 3D position, and 3D velocity.

The Random Forest algorithm \citep{breiman2001random} is a tree-based classification method that learns how to classify objects into different classes. Its founding components are the decision trees \citep{quinlan1993combining} that singly operate to make predictions on the single particle based on the associated input features $\vec{\theta}$. A Random Forest is thus a simple extension of the single operating decision tree, but it generally improves the performance of the classifier. Indeed, the advantage to constructing an ensemble of classifiers, where multiple trees fit random subsamples of the data, is that overfitting and instabilities in the data distributions may be mitigated by the averaged results of several trees. Therefore, we aim to design an adequate architecture of the Random Forest (e.g., number of trees, number of features to consider when looking for the best splits) to make the most accurate predictions.

\subsection{Data set and training phase}
We collected the data for the training set (later divided to perform cross-validation) and test set from the star particles in the simulated clusters and the output of the ICL-Subfind. The original set of simulated galaxy clusters is composed of 29 objects. We analyse the properties of these galaxy clusters to gather a fair sample of the cluster set. Besides all properties listed by Subfind (such as mass and radius), we determine the dynamical state of the host cluster (i.e. relaxed, disturbed, or intermediate), which is a good metric to derive the "thermalization" level of the particle phase-space distributions in a cluster. Particularly disturbed objects (e.g., after halo merging events) may not have a well-defined Maxwellian shape in the particle velocity distribution, thus complicating the fitting procedure used in splitting the stellar components. Thus, estimates of dynamical states are performed following the prescription described in \citet{neto2007statistics} based on two properties: the centre shift (identified as the distance between the minimum position of the gravitational potential $\mathbf{x}_\mathrm{min}$ and the centre of mass $\mathbf{x}_\mathrm{cm}$) and the fraction of mass in substructures $f_\mathrm{sub}$. We use the same threshold parameters as in \cite{biffi2016nature}. A halo is classified as relaxed if both the following conditions are satisfied:
\begin{equation}\label{eq:biffi+2016}
      \begin{cases}
         \delta r = || \mathbf{x}_\mathrm{min} - \mathbf{x}_\mathrm{cm} ||/R_{200}<0.07\\ \\
          f_\mathrm{sub} = \dfrac{M_\mathrm{tot,sub}}{M_\mathrm{tot}} <0.1
      \end{cases}       
\end{equation}
where $M_\mathrm{tot}$ is the total mass and $M_\mathrm{tot,sub}$ is the total mass in substructures within $R_{200}$. If neither is satisfied, then the cluster is classified as disturbed, while it is tagged as partially disturbed if only one of the above two criteria is not satisfied. After applying this classification to the 29 clusters at redshift $z=0$, we find 6 relaxed, 8 disturbed, and 15 intermediate cases. The physical properties taken into account for this selection (cluster mass, stellar mass of the central galaxy, and dynamical state) are listed in Table \ref{tab:simulated_clusters}. 
\par
Furthermore, we excluded a priori from our choice three clusters (that is, D7, D11, and D13) that did not reach convergence in the ICL-Subfind output while still retained as part of the test set to prove that our ML model can overcome challenging classifications for the traditional algorithm.
\par
Out of the 26 remaining clusters, we draw 10,000 star particles each randomly selected from five clusters (i.e., D3, D9, D10, D18, and D22) for a total of 50,000 particles. ICL and BCG are represented in this sample with proportions 65:35. We divide the training and test sets assigning 2/3 to the former and 1/3 to the second.

\begin{table*}
    \caption{A summary of the main characteristics of the simulated clusters used in the training and testing phases at $z=0$. We report the given cluster name, cluster mass \m200, the cluster radius \r200, the cluster orbital velocity $V_{200}$, the stellar mass in the main halo $M_{\star,gal}$ and the dynamical state. We add an asterisk to the clusters which are part of the training set.}
    \centering
        \setlength{\tabcolsep}{8pt}
        \begin{tabular}{c c c c c c }
            \\
            \thead{Cluster name}& \thead{\m200} & \thead{\r200} & \thead{$V_{200}$} & \thead{$M_{\star,gal}$} & \thead{Dynamical state}\\
             $\star = $ Training set & \thead{[$10^{15}$ \msun$/h$]} & \thead{ [Mpc$/h$]} & [km s$^{-1}$] & \thead{[$10^{10}$ \msun$/h$]} & \\
            \hline \\
            D1 & 1.26 & 1.76 & 1758 & 1402 & Intermediate\\
            D2 & 0.39 & 1.19 & 1188 & 399 & Intermediate  \\
            D3\astfootnote{\label{name}Training set} & 0.49 & 1.28 & 1282 & 599 &  Intermediate  \\
            D4 & 0.38 & 1.18 & 1176 & 348 & Disturbed  \\
            D5 & 0.14 & 0.84 & 840 & 177 & Relaxed  \\
            D6 & 1.12 & 1.69 & 1687 & 1077 & Intermediate  \\
            D7 & 1.10 & 1.68 & 1680 & 1220 & Intermediate  \\
            D8 & 1.24 & 1.74 & 1746 & 843 & Disturbed  \\
            D9\footref{name} & 0.10 & 0.76 & 756 & 125 & Relaxed  \\
            D10\footref{name} & 1.04 & 1.64 & 1647 & 1342 & Disturbed  \\
            D11 & 0.86 & 1.55 & 1547 & 1114 & Intermediate  \\
            D12 & 1.58 & 1.89 & 1895 & 1185 & Relaxed  \\
            D13 & 1.06 & 1.66 & 1658 & 1008 & Disturbed  \\
            D14 & 1.43 & 1.83 & 1832 & 1372 &  Intermediate  \\
            D15 & 1.36 & 1.80 & 1803 & 1290 & Intermediate  \\
            D16 & 2.74 & 2.28 & 2276 & 2013 & Disturbed  \\
            D17 & 1.43 & 1.84 & 1834 & 972 &  Intermediate  \\
            D18\footref{name} & 0.85 & 1.54 & 1542 & 1056 & Intermediate  \\
            D19 & 1.14 & 1.70 & 1703 & 1200 &  Intermediate  \\
            D20 & 1.43 & 1.83 & 1833 & 1298 & Intermediate  \\
            D21 & 1.18 & 1.72 & 1722 & 1174 & Relaxed  \\
            D22\footref{name} & 1.56 & 1.89 & 1887 & 1919 &  Relaxed  \\
            D23 & 1.06 & 1.66 & 1657 & 1030 & Disturbed  \\
            D24 & 1.09 & 1.67 & 1675 & 1433 & Intermediate  \\
            D25 & 0.79 & 1.51 & 1507 & 719 & Disturbed  \\
            D26 & 1.26 & 1.76 & 1757 & 1255 & Intermediate  \\
            D27 & 1.33 & 1.79 & 1789 & 1410 & Relaxed  \\
            D28 & 1.55 & 1.88 & 1881 & 1457 & Intermediate  \\
            D29 & 1.24 & 1.75 & 1749 & 1049 & Disturbed  \\
            \hline
        \end{tabular}
        \label{tab:simulated_clusters}
\end{table*}

\subsection{Input features}
\begin{figure*}
    \includegraphics[scale=0.45,angle=0.0]{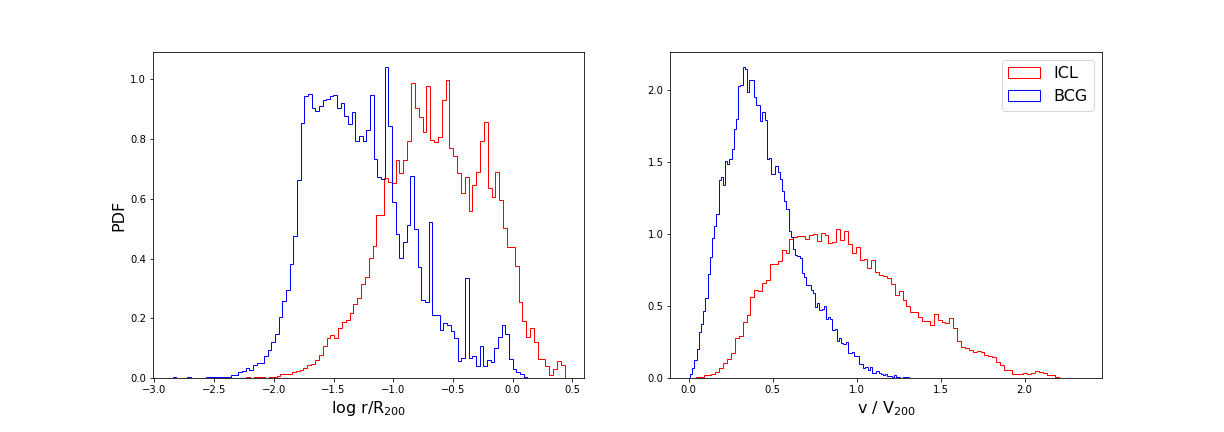} 
    \caption{Probability density functions of the input features associated to the ICL (red) and BCG (blue) in one of the clusters. \emph{Left:} Distribution of the logarithmic cluster distance over \r200. \emph{Right:} Stellar rest-frame velocities scaled by the virial circular velocity \v200.}
    \label{fig:features}
\end{figure*}
\begin{figure}
    \includegraphics[scale=0.55,angle=0.0]{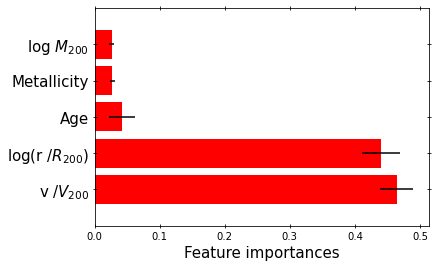}
    \caption{Importance of the input features including all tried input features to predict ICL and BCG components. From top to bottom: the logarithm of the cluster mass \m200, the stellar metallicity and age, the logarithmic cluster-distance over \r200, and the rest-frame velocities scaled by the virial circular velocity \v200. The black bars encode the standard deviation when sampling the importances from the trees in the forest.}
    \label{fig:Features_importances}
\end{figure}
The predicting power of a ML model heavily depends onto what extent the input features of the data set are representative of the classes one hopes to recover. Before ultimately evaluating our classifier's performance, we infer the combination of features that best match the two classes at hand. Starting from a large parameter space, we find that the particle cluster-centric distance and the module of the rest-frame particle velocity (with respect to the stellar centre of mass within \r200) offer most of the dynamical information needed to disentangle the two components, given that they closely relate to the particle energies employed in the ICL-Subfind unbinding procedure. Fig. \ref{fig:features} shows the probability density distributions of these two features drawn from one of the clusters in our simulations. To clarify the separation in the phase-space, we plot the histograms of the BCG (blue) and ICL (red) separately.

To obtain this result, we firstly examined the classifier's performance using a larger set of possible particle properties (e.g., distance from the cluster centre and velocity relative to it, cluster mass, age, metallicity, kinetic energy and potential energy) and recording the metric scores. Since the kinetic and potential energies directly correlate with velocity and distance, we only keep the latter two. Later, we selected different subsets of these properties to assess which combinations of them provide results which are consistent with the all-features case and its metric scores. Having established that once we include distance and velocity in the features space the performance would not further significantly improve by adding other features, we decided to keep a basic parameter space and excluded such additional features. Furthermore, since our training set is composed of subsamples from different halos, we perform a scaling of each of these quantities according to the cluster virial values. We normalize cluster-centric distances by \r200 and we take its logarithm to increase the separation between the two stellar components in the parameter space. Instead, the rest frame velocity is scaled by the circular virial velocity \v200, which reads
\begin{equation}
    V_{200}= \sqrt{\frac{G M_{200}}{R_{200}}} .
\end{equation}
Assuming that the particle distributions of distance and velocity in a cluster may not be fully generalized by only scaling for the corresponding virial quantities in our final set, we provide the cluster mass \m200 as an additional input parameter to the reference model. This supplementary information may help the classifier in choosing a label over another in case of the degeneracy of the other two parameters if the label is somewhat still dependent on the cluster size. 

Fig. \ref{fig:Features_importances} reports in increasing order the importance of all the features initially included in the trained ML model and the associated standard deviations with the black bars. Feature importances \citep[][]{scikit-learn} can be computed as the mean and standard deviation of the decrease in impurities summed for each tree in the forest. In the case of a classification problem, impurity is often defined as Gini impurity, which is a measurement of the likelihood of incorrect classification of a new instance if it were randomly classified according to the distribution of class labels in the data set. In other terms, this importance figure provides an understanding of how the ML classifier evaluates the input parameters relatively to each other. Although our choices were not driven by the feature importance analysis, we provide the results for our final set to point out an important aspect in the inclusion of the cluster mass as additional feature to the parameter space. The plot shows that distance and velocity are of similar importance, whereas the other parameters have seemingly lower values. This is not surprising since these two features are also those that ICL-Subfind uses to perform the classification. On the other hand, the mass of the cluster \m200 yields the lowest value among all features, provided that the training set is composed of only five clusters (and thus only five different input cluster masses (namely \m200 $= (0.49, 0.10, 1.04, 0.85, 1.56) \times 10^{15} M_{\odot}$), whereas the other features vary from particle to particle. However, we expect that this feature could play a more important role in larger cluster sets, which is the reason why we do not exclude it from our analysis.

\subsection{Classification performance}
To tune the (hyper)parameters of the classifier, we use a K-fold crossing validation, with K=5. This involves randomly splitting the training set into K complementary subsets and repeatedly training the model on K-1 subsets while validating the resultant estimator when applied to the remaining subset. Each time, the classifier is trained on different combinations of the (hyper)parameters to obtain unbiased estimates of the classifier’s average performance metrics and their uncertainty. The main parameters undergoing this search are the number of features to consider when splitting a tree, the depth of the trees, and the number of trees in the Random Forest ensemble.

Results from each of these cross-validated runs are analysed with performance metrics. Notice that the estimated ICL labels are (in most of the clusters) much larger than the BCG component. For our binary classification problem, we use recall (R), precision (P), and F-Score (FS) which are independent of the imbalance nature of the classification problem. These are defined as:
\begin{equation}
    \mathrm{R = \frac{TP}{TP+FN}} ; \qquad \mathrm{P = \frac{TP}{TP+FP}}; \qquad \mathrm{FS = 2\, \frac{P \times R}{P+R}}.
\end{equation}
In the above expressions, TP, FP, and FN are the numbers drawn from the predicted labels of true positives, false positives, and false negatives, respectively. In other words, recall expresses the rate at which the model correctly predicts the class of an object; precision measures the fraction of correctly classified objects over the total number of objects labelled with that class; finally, F-Score can be interpreted as the weighted average of the precision and recall, where it reaches its best value at 1 and worst at 0. We point out that in the text we will refer to "true" as the labels provided by ICL-Subfind. Clearly, they are not necessarily "true" in absolute terms, but they represent our reference answer to this classification problem. In fact, there might be cases where the metric score is lowered due to a difference in the labels between the two methods, rather than due to a poor recognition of the ML classifier in the dynamical properties of stars.

\section{Results}
\label{sec:results}
Based on the cross-validated parameter search, we find the algorithm to have a consistently good performance. Each class holds on this picture: the ICL shows P = 95 per cent, R = 92 per cent, and FS = 93 per cent, while the BCG class presents P = 78 per cent, R = 85 per cent and FS = 81 per cent. We remind that these scores are valid for the specific subhalo finder used, Subfind, and larger differences could be found when employing other algorithms. 

Besides its high accuracy, one of the benefits of employing the ML algorithm to classify star particles is its efficiency and speed up with respect to ICL-Subfind. Provided that the latter not only performs the star particle classification but also identifies the substructures in the FoF catalogue, in cases where one already has the subhalo identification for a given cluster (a standard procedure in state-of-the-art simulations of galaxy clusters with the aim of analysing galaxy populations), it is possible to bypass this step and directly obtain the labels for the stars in the main halo. Skipping this unnecessary operation can be crucial in saving run-time for large simulations whereby the subhalo identification can take several hours on different cores. For this reason, it is not straightforward to fully quantify the computational advantage of employing one technique over the other, unless one only needs the star particle classification having done the subhalo identification in previous steps. Taking this into account, a rough estimate of the run time of ICL-Subfind restricted to the sole stellar classification for a cluster at our reference resolution (considering the operational time spent by the traditional Subfind to detect subhalos) gives a speed-up by a factor of about 100. On the other hand, considering both procedures, the savings in time add up to an order of $10^5$. Furthermore, increasing the numerical resolution (and therefore the number of particles in a simulation) might entail a severe increase in the run-time, whereas no significant difference involves the ML classifier. In conclusion, we recommend the use of the ML model in cases where simulations have already undergone a subhalo identification procedure. However, further examinations are required to assess possible differences when other halo finders are used, since we only examined the results by Subfind. 

In the next section, we would like to quantitatively assess the resolving power of the model compared to the traditional ICL-Subfind. With this in mind, we select a random cluster in our simulation (its properties are summarised in Table \ref{tab:simulated_clusters} under the name D5), and examine the differences between ML and ICL-Subfind in the feature distributions, phase-space profiles, and mass-weighted maps. The last part of this section will provide a more general overview of the results for the entire cluster data set.

\subsection{Testing on a single simulated cluster}
\begin{figure*}
    \includegraphics[scale=0.45,angle=0.0]{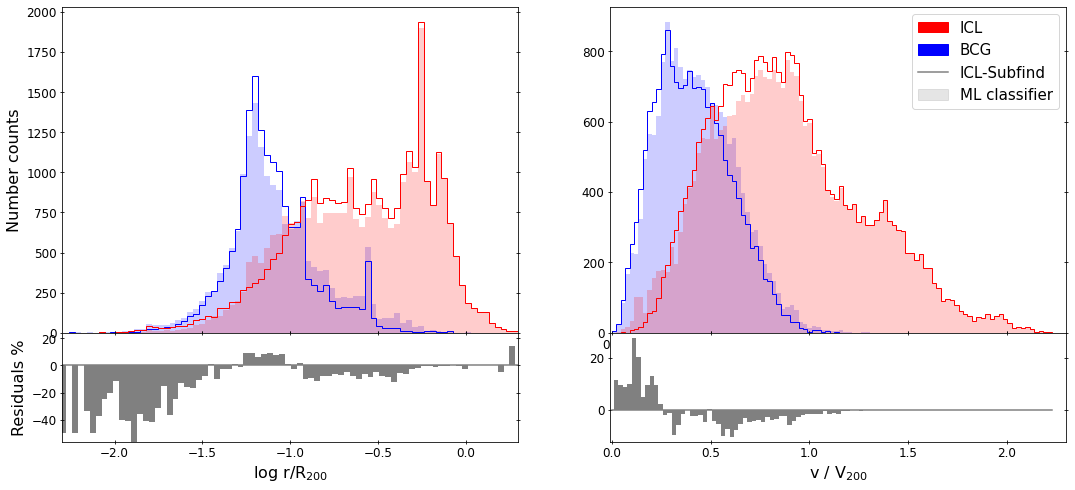}
    \caption{\emph{Top:} Distribution of the input features drawn from the simulated cluster. In each panel, we plot the predicted (bars) and true (line) number counts of both ICL (red) and BCG (blue) associated with the star particles. More in detail we have (from left to right) the distributions of the logarithmic cluster-centric distance normalized for \r200 and the stellar rest-frame velocities normalized for \v200. \emph{Bottom:} Percentage residuals measured between the true and predicted counts in each bin. }
    \label{fig:D5_features-1x}
\end{figure*}

The first comparison is between the input features of the star particles. Fig. \ref{fig:D5_features-1x} illustrates the comparison between the number counts of the particles' logarithmic cluster-centric distance (left panel) and their velocity distribution (right panel), both normalised by their virial value. True and predicted labels are marked with a contouring line and area, respectively, for both ICL (red) and BCG (blue) stars. In the bottom panels, we show the percentage residuals between the true and predicted labels over the total number of star particles in each bin to estimate where the results are most different. This definition of residuals (in absolute value) is the same whether we consider ICL or BCG stars, given that it simply represents the excess of one class over the other, normalised by the number of particles in each bin. For this reason, we consistently choose throughout the paper to represent the BCG excess (or deficiency, depending on the sign) of the ML prediction with respect to the model. The ICL percentage residuals will then simply correspond to the opposite number. 

In both cases, we observe a generally good agreement within each predicted subgroup and its true distribution. The left panel confirms the presence of a bulk structure in the inner region, which corresponds to the BCG and the ICL, a more diffuse component that extends beyond \r200. The largest differences are found in the inner core of the BCG (up to 40 per cent), but they are mostly due to the low number of star particles in these bins. As we move towards the outskirts of the BCG, the distribution residuals span values around 10 per cent, which represents a more consistent estimate of the errors in the classification process at these distances. We usually find this transition region to be carrying most of the uncertainty in the labelling of stars in all clusters, as it will be illustrated in the next section. In this regard, we expect both algorithms to carry uncertainties, which will sum up at the expense of the ML metric scores. In other words, the low metric scores for the ML algorithm are due to a different labelling with respect to ICL-Subfind, which in turn is not necessarily always correct. The classification process suffers from finding dynamically similar particles in the ICL and BCG components that populate regions far from the centre, thus decreasing the precision of the algorithm at these distances.

As for the velocity distribution, we confirm the presence of the two peaks which can be fit by the double Maxwellian. On this point, we highlight the closeness of the velocity distributions among the two methods, which is already a good index of the accuracy of the ML algorithm, given that this result is obtained without the need of an explicit fit. 
\begin{figure*}
    \includegraphics[scale=0.5,angle=0.0]{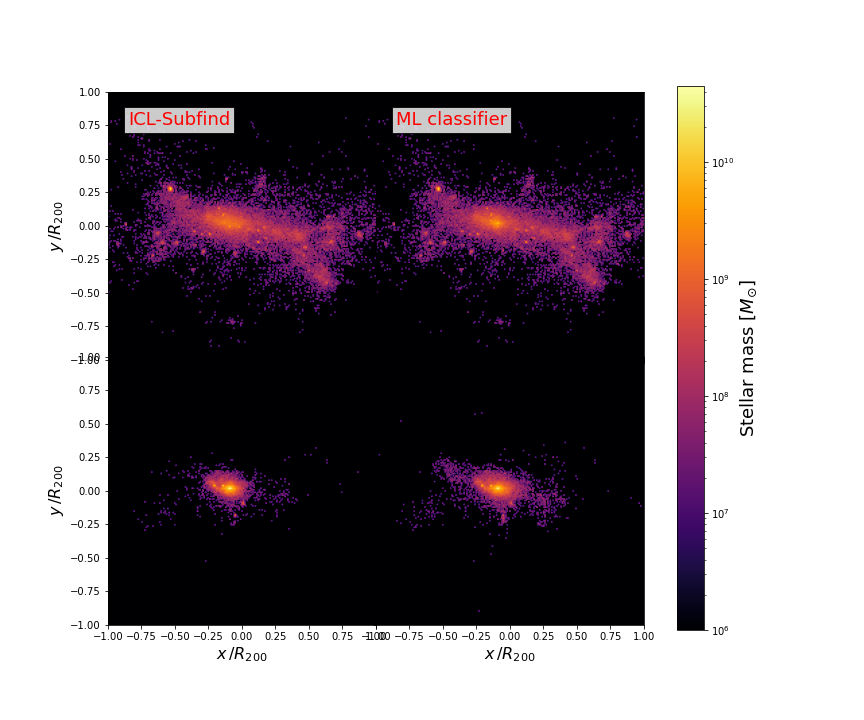}
    \caption{Mass-weighted maps of the stellar components (ICL in the top panels, BCG in the bottom ones) in the same cluster as in Fig. \ref{fig:D5_features-1x}. The left panels report the results from the stellar division provided by the traditional method with the ICL-Subfind. The right panels show the mass-weighted maps for the stellar components identified with the predicted labels.}
    \label{fig:D5_map-1x}
\end{figure*}
\begin{figure}
    \includegraphics[scale=0.45,angle=0.0]{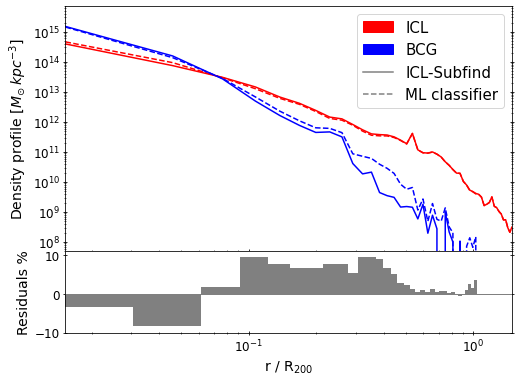}
    \caption{\emph{Top:} Density profiles of the BCG (blue) and ICL (red) in the selected cluster. The dashed lines are the profiles computed with the predicted labels, while solid lines report the profiles computed with the true labels. \emph{Bottom:}  Residuals (in per cent) between the histograms of the labels from the ICL-Subfind and the ML classifier.}
    \label{fig:D5_density-1x}
\end{figure}
Fig. \ref{fig:D5_map-1x} shows the mass-weighted maps of the distinct stellar components in the cluster under study. ICL in the top panels and BCG in the bottom. The panels on the left are the results from the traditional method, while on the right we illustrate the maps when employing the labels from the ML algorithm. The results are remarkably similar and the differences are mainly explained on the BCG outskirts. This can be better appreciated comparing the 3D density profiles of both ICL (red) and BCG (blue) as computed with the true labelled stars (solid line) or with the predicted labels (dashed line) in Fig. \ref{fig:D5_density-1x}.

Another aspect to consider in evaluating the performance of the algorithm is to study the ICL fraction predicted by the ICL-Subfind and that from the ML scheme. We define the ICL fraction $f_\mathrm{ICL}$ as the number of ICL particles over the total number of stars in the main halo. The ML model yields values that are also consistent with those predicted by the traditional algorithm: in this particular cluster, we measure 0.63 using the ML method as opposed to 0.62 for the traditional case when considering the ICL fraction over the stars of the main halo. We recall that this value shall not be directly compared to the observational results.

\subsection{Testing on a simulated cluster population}
In the previous section, we showed that our ML-based algorithm to separate stellar ICL and BCG populations is a robust classifier in the case of a single test cluster. We can take a step further and apply the trained classifier over all clusters not part of the training set (24 in our simulations) to present a few results which are worth discussing. 
\begin{figure*}
    \includegraphics[scale=0.6,angle=0.0]{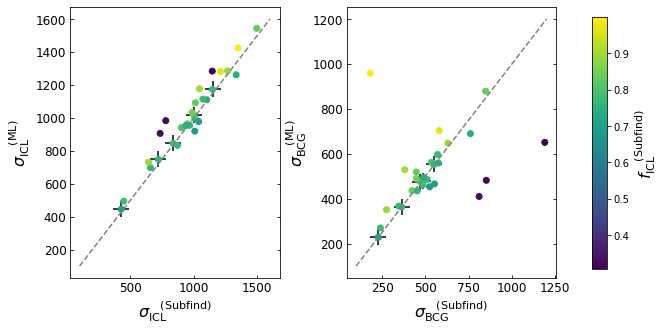}
    \caption{Comparison of the velocity dispersions derived from the fit of the double Maxwellian in Eq. \ref{eq:double_maxw} in the stellar velocity distribution labelled by the ML classifier and the ICL-Subfind. Each point represents the result from the component of a single cluster, coloured according to the ICL fraction estimated in each cluster by the ICL-Subfind. Furthermore, we mark the clusters from the training set with a cross. The dashed grey line is the reference line to a 1:1 relation. \emph{Left: } The velocity dispersions in the ICL component. \emph{Right: } The velocity dispersions in the BCG component. }
    \label{fig:tab2}
\end{figure*}
\par
Fig. \ref{fig:tab2} shows the comparison between the velocity dispersions derived from the fit of the double Maxwellian in Eq. \ref{eq:double_maxw} to the stellar velocity distribution as labelled by the ICL-Subfind and ML classifications in all our clusters. In the left panel, we plot the ICL velocity dispersions; in the right panel, we present the BCG velocity dispersion. Each point marks a single cluster colour-coded for the ICL fraction as given by the ICL-Subfind sample. We report all 29 clusters, including those not converged in the ICL-Subfind (the three isolated points with low ICL fractions), while we mark the training clusters with a cross for clarity. The dashed grey line marks to reference the 1:1 relation. Quite remarkably the relationship between ICL velocity dispersion from ICL-Subfind and from the ML classifier (left panel of Fig. \ref{fig:tab2}) shows a small scatter around this relation, with the latter being on average 4 per cent higher. This difference increases slightly ($10-20$ per cent) in correspondence of the three groups with the lowest ICL fractions. On the contrary, for the BCG velocity dispersions (right panel of Fig. \ref{fig:tab2}), we observe a significant colour gradient, perpendicular to the reference line. Clusters hosting a larger BCG stellar fraction and a correspondingly lower ICL fraction, assigned by ICL-Subfind, have a higher BCG velocity dispersion, and vice versa. This large difference between the two methods can be traced back to the extreme values of the ICL fraction in the ICL-Subfind predictions, which do not occur in the ML case (specifically stretching for all clusters only within the range 0.60 -- 0.80). Understandably, the larger/smaller the virialized system (BCG, in this case), the higher/lower the velocity dispersion. 
\par
Table \ref{tab:compar_ml-subfind} summarizes several of these results for all clusters: we compare the fits of the double Maxwellian, the ICL fractions for both the ICL-Subfind and ML algorithms and the performance scores. The latter are reported for both the single classes and the means weighted with the number of the two components. As previously mentioned, the scores employed to assess the quality of the predictions by the ML classifier with respect to ICL-Subfind are precision P, recall R, and F-Score FS. We find that BCGs usually have high P and lower R scores, which expresses the capability of the ML algorithm to be generally correct when labelling BCG stars, although not returning the entire set of BCG particles compared to the true set. On the contrary, ICL has usually most of the particles assigned, thus yielding the opposite situation. However, FS is high in most cases (a few pathological cases will be discussed in the next paragraph). A more explicit report of the classification score is provided by the corresponding weighted means in the last columns. Relaxed clusters, which typically have a well-defined double Maxwellian velocity distribution, reach a mean FS score of $80-90$ per cent. This is the case for 15 clusters in the test set. On the other hand, we observe a few clusters receiving consistently low scores in the classifier's metrics (i.e., FS is below $40$ per cent) while their physical features are generally inconsistent with the results obtained by ICL-Subfind. This is the case for D7, D11, and D13 for which, as mentioned before, ICL-Subfind did not reach convergence. For these clusters, the ICL fraction is a few per cent and, indeed, there is no separation of the components. On the contrary, we believe that our ML-based method overcomes these situations by correctly identifying two dynamically distinct components, with the BCG component having a markedly smaller velocity dispersion than the ICL one. Quite interestingly we notice that a few central galaxies in our sample show tidal shell features \citep[see][for a review]{ebrova2013shell} once we separate the BCG and ICL. This is the case for five of our clusters, two of which are part of the clusters that have not converged in the ICL-Subfind procedure. These shell-like features in the stellar distribution could be linked to past tidal shocks, associated to recent merger events. We briefly discuss this observed feature from Fig. \ref{fig:D7_maps} in Appendix \ref{Appendix}.
\par
\begin{figure}
    \centering
    \includegraphics[scale=0.45,angle=0.0]{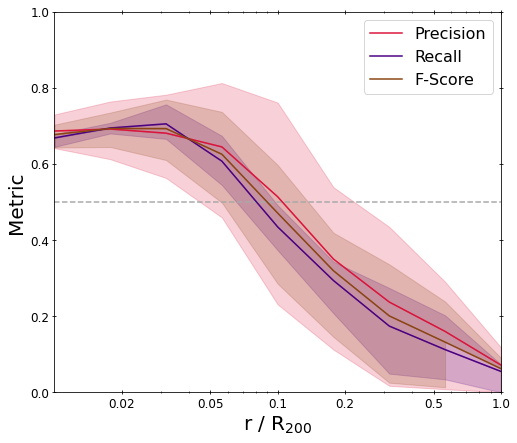}
    \caption{Stack of the metric score profiles relative to the BCG particle classification in the entire cluster set. The profiles are reported as a function of the cluster distance normalized by \r200. Solid lines refer to P (red), R (blue) and FS (brown). The dashed line marks the value 0.5 in the y-axis.  }
    \label{fig:metric_bcg}
\end{figure}

A common trait in the testing set is the discrepancy between predicted and true classes on the outskirts of the BCGs, where the distinct dynamical behaviours of the star particles are generally harder to discern. 
Fig. \ref{fig:metric_bcg} investigates this flaw in the performance by showing the BCG metric scores as a function of the radial distance from the centre of the subhalo. We stacked the metric score profiles for the BCG labels (P in red, R in blue, and FS in brown) of the entire cluster set to tentatively describe the expected accuracy. Shaded bands display the standard deviation given by the intrinsic distribution, while the dashed grey line marks the value 0.5 on the y-axis, below which the rate of incorrect labelling is more than one in two in the (predicted or true) BCG set. We notice that the metric score in the centre is high ($>0.7$), while from $\sim 0.1 R_{200}$ ($\sim 250$ kpc) it declines very rapidly. Here, far from the central region, we expect differences between the ICL and BCG properties to become less sharp, since BCG particles will have larger entropies compared to the centre, spanning a phase space very similar to that occupied by the ICL.

\begin{figure}
\centering
    \includegraphics[scale=0.35,angle=0.0]{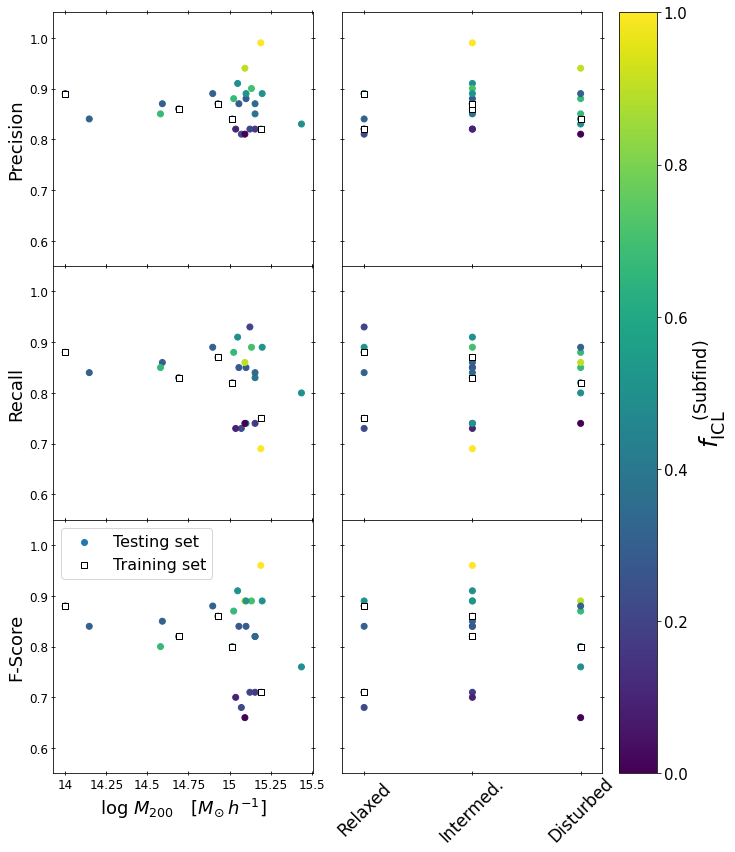}
    \caption{Scatter plots of the (weighted) mean metric scores (P, R and FS from top to bottom panel) as a function of log cluster mass (left panels) and dynamical state (right panels) for all the clusters under study. We mark the pure testing set with coloured points, while the clusters in the training set are recorded with empty squares. The colour legend follows the ICL fraction estimated by ICL-Subfind.} The y-axis is limited to exclude D7, D11 and D13 having very low metric scores.
    \label{fig:metric_scores_withfICL}
\end{figure}

Furthermore, we evaluated the ML model metric scores for the fairly resolved sample of clusters (that is, excluding D7, D11, and D13) as a function of the mass of the cluster (left panels) and the dynamical state (right panels) in Fig. \ref{fig:metric_scores_withfICL}. \citet{dolag2010dynamical} refer to potential uncertainties in the classification process for low-mass clusters owing to the difficulty in disentangling BCG and ICL in the velocity distributions of low-mass halos, where the two Maxwellians cannot be easily discerned. On the other hand, the dynamical state and recent merger history can strongly impact the physical conditions of the stellar components in the inner regions. In the analyzed simulated clusters, these merging events of massive orbiting halos with the BCG produce strangely-shaped halos with non-thermalized velocity distributions, or strongly nonspherical symmetry in the star particle distribution, which can be due either to a peculiar halo formation history or to an incorrect group identification by Subfind in the main halo. In this case, both ICL-Subfind and the ML algorithm may encounter difficulties in properly separating the two components: size estimates of the BCG are extremely sensitive to these non-thermalized distributions of star particles. However, Fig. \ref{fig:metric_scores_withfICL} shows no significant correlation with either of the cluster properties. We plot from top to bottom the P, R, and FS for the clusters in both training (empty squares) and testing set (filled dots). The latter are colour-coded according to the ICL fraction computed with the labels from ICL-Subfind. There seems to be a mild correlation between P and $f_\textrm{ICL}$, and thus in the FS, however this shall be verified with a larger sample of clusters.

\subsection{Testing the robustness of the classifier}
So far, both training and testing have been described for a given set of simulated galaxy clusters that, despite their specific history of formation, share many similarities: the same subresolution model for star formation and feedback, the same numerical resolution, and the same redshift. In the effort to understand the real range of possible applications of our classifier -- compared to what is originally obtained with ICL-Subfind -- we apply our model to other simulated clusters which differ from the original cluster set in different ways. We decided to re-simulate two out of the 29 clusters (i.e., D1 and D2) in different conditions and we discuss the outcomes of these analyses in the following sections. Despite the limited statistics, we expect to obtain useful insights into the predictive power of our method from these tests. 
\subsubsection{Changing redshifts}
\begin{figure*}
    \includegraphics[scale=0.45,angle=0.0]{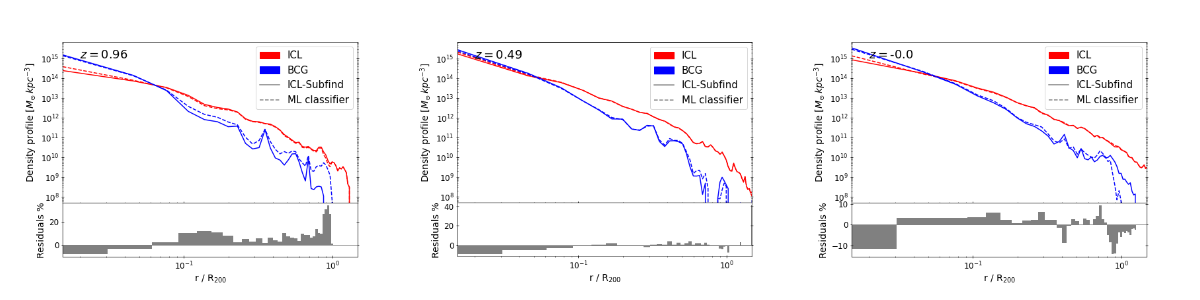}
    \caption{Density profiles of the stellar content in D2 taken at three different snapshots: from left to right the correspondent redshifts are $z\simeq1$, $z\simeq0.5$ and $z=0$. For each of these plots, we show two panels. In the top panel, we report the profiles normalized at \r200 of both ICL (red) and BCG (blue) identified by ICL-Subfind (solid line) and the ML classifier (dashed line). The lower panel shows the residuals (in per cent,  normalized by the number of star particles in each bin) between the BCG true and predicted labels.}
    \label{fig:D2_redshift}
\end{figure*}

We analyse here the behaviour of the ML classifier in the same simulation at redshifts different from that of $z=0$, at which the method has been trained. We point out that both traditional and ML methods rely on the underlying physical assumption that the two stellar components can be described by a double Maxwellian early enough to label the stars consistently as for $z=0$. This is not necessarily true if the stellar populations are still forming or evolving significantly. For this reason, we analyse our simulated clusters at two different redshifts ($z\simeq0.5$ and $z\simeq1$) at which most of the BCG stellar mass is already in place \citep[][]{ragone2018bcg}. Fig. \ref{fig:D2_redshift} illustrates the evolution of the stellar density profiles of D2 (from left to right panel: the redshifts are $z\simeq1$, $z\simeq0.5$, and $z=0$). The legend is as before colour-coded for the stellar type (ICL in red and BCG in blue), ML results are described by a dashed line, while solid marks the ICL-Subfind output. Results and performance scores are largely consistent with what we found in the case of $z=0$: no significant systematics can be detected between the ML classifier and the traditional labelling and overall the distributions are recovered. Unsurprisingly, these are slightly better in the case of $z=0.5$, rather than in the case of higher redshift. 

\subsubsection{Changing numerical resolution}
\begin{figure}
\centering
    \includegraphics[scale=0.45,angle=0.0]{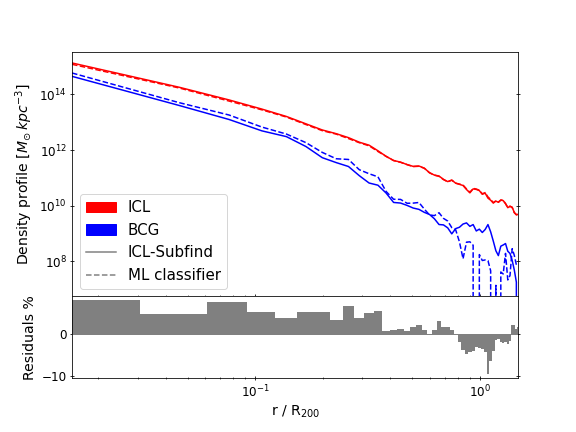}
    \caption{Density profiles of the stellar content in D2 simulated at higher resolution. In the top panel, we report the profiles normalized at \r200 of both ICL (red) and BCG (blue) identified by ICL-Subfind (solid line), the ML classifier (dashed line). The lower panel shows the residuals (in per cent,  normalized by the number of star particles in each bin) between the BCG true and predicted labels. Differences account for a per-cent level only between the profiles.}
    \label{fig:D2_MR_density91}
\end{figure}

An important step in understanding the quality of our predictions is to estimate the effect of numerical resolution. This can be performed by examining the results of our classifier, trained on a cluster set at a given resolution when applied to a set at a higher resolution. Increasing the resolution in a simulation improves the description of lower-mass systems and small-scale features. In turn, this could affect the probability distribution function at the centre of the clusters. In our resolution tests, we decreased the particle mass by a factor of 3 with respect to the reference simulation set, yielding $m_\textrm{DM} = 2.5 \times 10^8$ h$^{-1}$ M$_{\odot}$ and the initial mass of the gas particle $m_\textrm{gas} = 1.1 \times 10^8$ h$^{-1}$ M$_{\odot}$ for two clusters. The performance scores are found to be quite high (e.g., P$>0.75$, R$>0.80$, FS$>0.78$), with the stellar density profiles from ICL-Subfind and our method agreeing to per-cent level, as shown in Fig. \ref{fig:D2_MR_density91}. Further tests were performed for simulations at even higher spatial resolutions (also increasing our fiducial softening lengths of 3 times, as in \citealt{bassini2020dianoga}) giving similar high-performance scores, but they are not shown here. Therefore, our ML classifier seems to be robust when applied to simulations whose resolution is higher than that of the training set. 

\subsubsection{Changing the feedback model}
\begin{figure}
\centering
    \includegraphics[scale=0.45,angle=0.0]{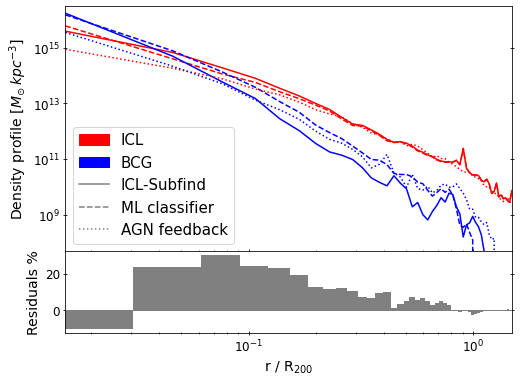}
    \caption{Density profiles of the stellar content in D2 simulated with no AGN feedback scheme. In the top panel, we report the profiles normalized at \r200 of both ICL (red) and BCG (blue) identified by ICL-Subfind (solid line), the ML classifier (dashed line). Additionally, we overplot the BCG and ICL profiles from the same cluster but including the AGN feedback model (dotted line). The lower panel shows the residuals (in per cent,  normalized by the number of star particles in each bin) between the BCG true and predicted labels. }
    \label{fig:D2_LR_CSF-density}
\end{figure}
To further test the robustness of our ML classifier, we applied it to simulations having the same resolution of the training set, but not including AGN feedback. Obviously, this is an extreme (possibly non-physical) scenario held with the only illustrative purpose of examining the consequences on the classifier performance facing underlying different physical conditions with respect to the training set. AGN feedback regulates star formation in massive galaxies, particularly impacting BCG masses \citep[e.g.][]{ragone2013brightest}, thus this test allows us to examine the ML robustness in a conservative regime of exceedingly massive galaxies.
\begin{figure*}
\centering
    \includegraphics[scale=0.4,angle=0.0]{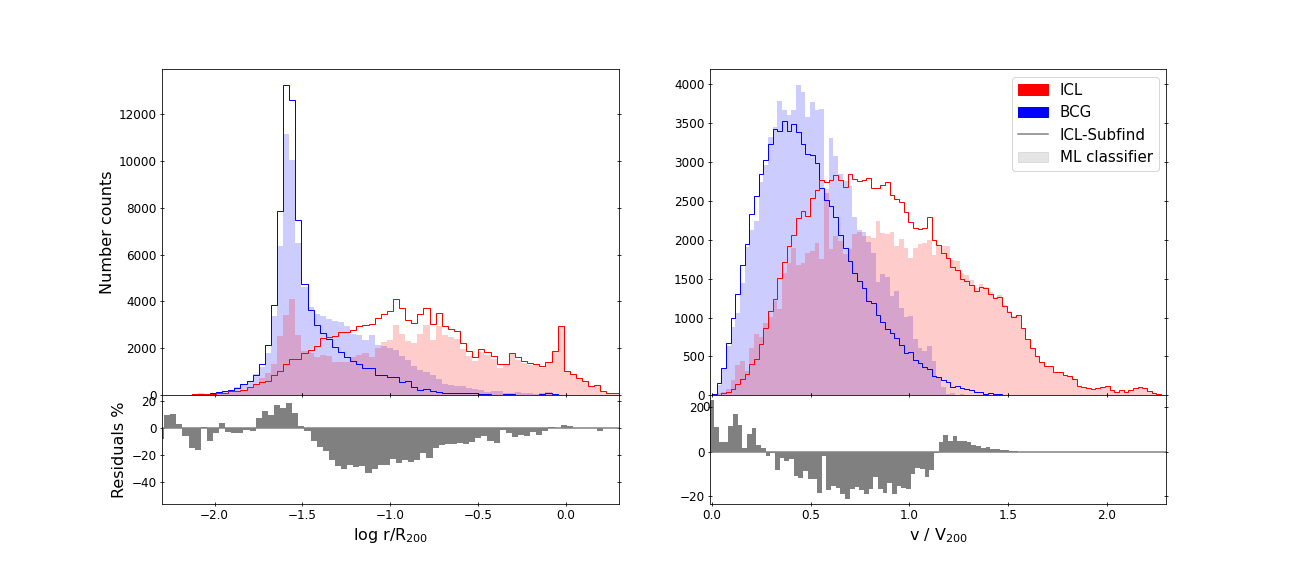}
    \caption{Histograms of the distributions of the features of D2 simulated with no AGN feedback scheme. In the top row, we report the number count distributions of (from left to right): the logarithmic cluster-centric distance over \r200 and the particle rest-frame velocity over \v200. The lower row shows the residuals (in per cent,  normalized by the number of star particles in each bin) between the BCG true and predicted labels.}
    \label{fig:D2_LR_CSF-features}
\end{figure*}
Our analysis shows the presence of a massive BCG at the centre of the halo in both ICL-Subfind (solid) and ML (dashed line), as can be seen in Fig. \ref{fig:D2_LR_CSF-density}. We notice that this causes a steepening in the density profiles with respect to what was predicted in presence of the AGN feedback mechanism (dotted lines), thus increasing the concentration of the stellar halo. In the latter, the differences in the density profiles account for up to only 3 per cent between the labels of the ML classifier and ICL-Subfind, so we decided to plot only one of them. Quite interestingly, we find that the ICL-Subfind and ML model-predicted profiles match better in the innermost part, where we would expect most of the AGN feedback to have a substantial effect, rather than in the BCG outskirts: here, the differences reach up to 25-30 per cent between the two runs. A closer look at the feature distributions in Fig. \ref{fig:D2_LR_CSF-features} shows that the phase space also shows some inconsistencies between the two methods. Yet, we are able to fit a double Maxwellian distribution to the star particle velocities reasonably close to the original one. 

Given the good quality of our analysis to this point, we pushed our investigation to simulations when both effects (numerical resolution and subgrid physics) are different relative to the reference simulation set. Although we do not display any of the profiles, we can confirm that the performance scores for this case are in line with the previous ones, demonstrating that the ML classifier results are robust across small changes in numerical resolution, redshift (at least in the late Universe, within $z\leq 1$) and physical subgrid models.

\subsection{Phase-space structure}
To further investigate the accuracy and robustness of these methods, we examine several quantities that generally describe the phase-space structure of galaxy clusters \citep[e.g.][]{marini2021phase}, in particular discussing it in terms of the stellar density profile $\rho(r)$, stellar velocity dispersion profile $\sigma(r)$ and phase-space density profile $Q(r) = \rho(r)/\sigma^3(r)$. We expect these quantities to provide insights into the robustness of these two methods based on the star particle distribution and dynamical information. Our ultimate goal is to detect major and/or systematic differences within the ICL and BCG subgroups as given by the two methods.

\begin{figure}
\centering
    \includegraphics[scale=0.45,angle=0.0]{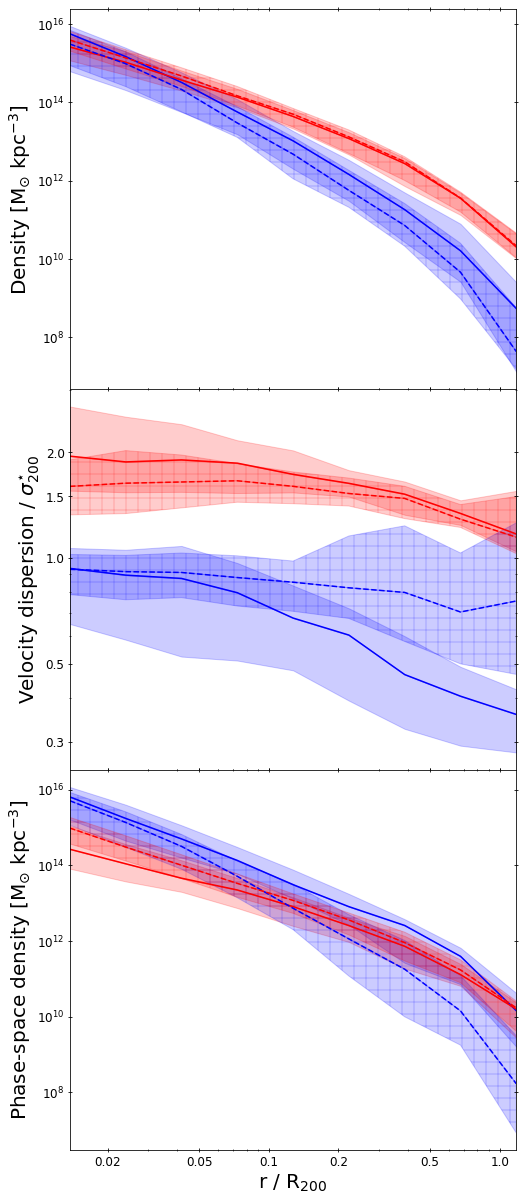}
    \caption{Density (top), velocity dispersion normalized for the velocity dispersion of the stars within \r200 (central), and phase-space density (bottom) profiles of the star particles in the BCG (blue) and ICL (red). Results from the ICL-Subfind labels are given with solid lines, on the other hand, the dashed lines mark the profiles extracted from the ML classified labels. In each panel, we report the median profile of each method (dark solid line) and uncertainty given by the intrinsic standard deviation with the shaded coloured band. }
    \label{fig:phase-space}
\end{figure}

Our main findings are illustrated in Fig. \ref{fig:phase-space}. From the top to bottom panel, we report the density, velocity dispersion, and phase-space density profiles of the star particles labelled as BCG (blue) and ICL (red) by the ICL-Subfind (solid lines) or the ML classifier (dashed lines). The radial distance is scaled by the virial radius \r200, to properly account for the different cluster sizes when stacking. The shaded bands represent the intrinsic scatter within the sample of clusters, computed as the standard deviation.

The BCG and ICL density profiles predicted by the two methods do not show significant differences. Therefore, we can provide an estimate of the transition radius, defined as the cluster-centric distance at which the ICL distribution starts dominating the stellar component. Given our different approach based on dynamical criteria rather than from a fitted profile, we are able to provide an independent comparison with the values proposed in the literature. The cluster set yields an average transition radius of about $90$ kpc (corresponding to $0.04$ \r200) which is in agreement with both theoretical and observational findings \citep[e.g.][]{gonzalez2021discovery,contini2022transition}. On the other hand, in the velocity dispersion profiles we observe a systematic difference in the profiles for large radii, even beyond the expected transition radius. The ML classifier tends to prefer a dynamically hotter BCG component compared to the output of ICL-Subfind. This is particularly highlighted in the central panel, where we compare the velocity dispersion profiles $\sigma(r)$ scaled by the velocity dispersion of the stars within the virial radius $\sigma^{\star}_{200}$ to correctly stack the distinct clusters. Velocity dispersions traced by the ICL stars are generally higher than those of the BCG stars at all radii, consistently with the results shown in Sect. \ref{sec:simulations} on the velocity distributions: we find that ICL profiles are consistent within 1$\sigma$ in the two methods. Conversely, BCG profiles are similar at the centre, while at large radii the ML classifier tends to include particles in the BCG with higher velocity dispersion than in the ICL-Subfind case. This results in an almost flat velocity dispersion profile. Indeed, these differences are present at large radii, where the assignment of star particles to one of the two components is less obvious, as seen in Fig. \ref{fig:metric_bcg}.

An extra step can be taken by evaluating the phase-space density which combines the density and velocity dispersion profiles to investigate the phase-space structure of halos. Both numerical and observational \citep[just to name a few][]{taylor2001phase,dehnen2005dynamical, faltenbacher2005supersonic, biviano2013clash,biviano2016dynamics,marini2021phase} studies have demonstrated that the profiles of phase-space density (or equivalently, of the pseudo-entropy $S(r) = Q(r)^{-2/3}$) have a power-law dependence on the cluster-centric radius, with a rather small scatter. In this context, \citet{marini2021phase} investigated the pseudo-entropy profile of different tracers in a set of simulated clusters, including the star particles as tracers of the phase-space structure of the cluster, and demonstrated that, while BCG and ICL components separately do not produce accurate power laws for the phase-space density, the power-law profile is instead recovered when analysing together the star particles of such two components. Once again, we see that the largest differences are in the BCG outskirts.

\section{Conclusions}
\label{sec:conclusions}
We presented a robust and efficient method to label stars in the main halo of simulated galaxy clusters as IntraCluster Light (ICL) or bound to the Brightest Cluster Galaxy (BCG) based on a Random Forest classifier. The classification model is trained, cross-validated, and tested on 29 galaxy clusters simulated with cosmological hydrodynamical simulations, reaching a high level of precision. This Machine Learning (ML) method is based on a more traditional algorithm, which we call ICL-Subfind, fully described in \citet{dolag2010dynamical}. In that paper, the authors showed the existence of two dynamically distinct components in the stellar population of simulated galaxy clusters (associated with the main halo), which are identified because their velocity distributions can be fitted by a double Maxwellian distribution. Including this information in a gravitational unbinding procedure yields a spatial separation of the ICL and BCG stellar components in the central subhalo of simulated galaxy clusters. The subset of stars with the largest velocity dispersion is associated with the hottest stellar component, the ICL, while the other is assumed to be bound to the central galaxy or BCG. By applying the ICL-Subfind algorithm to the star particles in the 29 simulated clusters of the DIANOGA set, we obtain several data sets which we can use to fit a supervised model, intending to obtain consistent results with the traditional ICL-Subfind method, but far more efficiently. 

To construct the classifier, we find the combination of input features that proves to best represent the two classes (labels) we are seeking, which for our specific problem are the cluster mass \m 200, the cluster-centric distance of particles normalised by \r200, and the velocity of the rest frame of particles normalised to \v 200. We use randomly selected subgroups of particles from five clusters to train and cross-validate the classifier, while the rest of the clusters are employed to further test the predicted generalisation of the algorithm. 

Our results can be summarised as follows.
\begin{itemize}
    \item Our classification method agrees to a high degree of precision with the true labels (i.e., ICL-Subfind) of the two stellar components in the cluster population. We find the existence of a central, more gravitationally bound, stellar bulk, the BCG, which is disentangled from the more diffuse ICL, that instead extends to larger distances. The fraction of ICL is also consistent with that found by ICL-Subfind and is generally higher (by a factor of about 3) than that associated with the BCG. Nevertheless, we stress that the ICL mass fraction found here shall not be regarded as immediately comparable to observations, where the separation between ICL and BCG is not performed in a dynamical analysis. 
    \item We show that the metric scores relative to the BCG decrease steadily beyond 0.1\r200, as shown in Fig.\ref{fig:metric_bcg}, in turn affecting the density and dynamical profiles. We shall recall that at these distances, both algorithms carry uncertainties to a certain extent in labelling star particles. In other words, a lower metric score at these scales may be also due to the composite effect of wrong labelling from both algorithms.
    \item Our model proved to be robust to changes in the numerical resolution, across different redshifts (up to $z=1$) and with the exclusion of an AGN feedback model in the simulation. These results support the use of this method in various heterogeneous situations. 
    It is unclear to what extent this model becomes unresponsive, and thus we conclude that further analyses are required for systems significantly different from those tested.
    \item Additionally, we examined the dependence of the performance of the ML algorithm on the mass of the cluster and the dynamical state. Both effects may play a role in changes of the phase-space structure of a cluster and, in turn, affect the performance. However, we did not detect any significant correlation with either of these properties.
    \item Finally, to gain more insight into the quality of the dynamical information retrieved by this method, we studied the phase-space structure of the two stellar components after applying both classifiers to the simulated cluster set in Fig. \ref{fig:phase-space}. We estimate the transition radius between the BCG-dominated and ICL-dominated regions at around 0.04 \r200 (corresponding to a physical scale of 90 kpc), in line with previous observational measurements and theoretical predictions. This is particularly interesting in light of the different assumptions used to determine its value. We identify the BCG outskirts to be the most critical region (i.e., beyond $0.1$\r200, corresponding to a physical distance of roughly 250 kpc) and far more prone to uncertainties in the classification process. This is due the co-existence of the two stellar components whose physical properties overlap. 
\end{itemize}

In conclusion, this method proved to be reliable and faster than the traditional method to identify ICL and BCG in the main halo of simulated galaxy clusters. Although it does not provide a new methodology for detecting ICL, it offers a robust tool to further investigate the dynamical characteristics of ICL compared to the traditional method. As a final remark, we shall refrain from claiming that this classifier will perform at this level of accuracy for simulations including significantly different astrophysical models, unless these are included in the original training set. To remain a competitive alternative, when applied to very different simulations, one should resort to more advanced models, trained on vaster sets of simulations, which will need to include different cosmological and astrophysical scenarios for structure evolution. Ultimately, a dynamical analysis of the ICL should be regarded as an attempt to determine its physical properties and its origin to gain insights into the evolution of clusters and their stellar components.
\onecolumn
\begin{landscape}
        \begin{longtable}{l l l l l l l l |l l  l | l l l | l p{0.5cm} l }
            Name& Converged & \dispbcg$^\textrm{(ICL-S)}$ & \dispicl$^\textrm{(ICL-S)}$ & \dispbcg$^\textrm{(ML)}$ & \dispicl$^\textrm{(ML)}$ & f$_\textrm{ICL}^\textrm{\, (ICL-S)}$ & f$_\textrm{ICL}^\textrm{\,(ML)}$  &  & BCG & & &ICL & & & WMean & \\
            & & [km s$^{-1}$] & [km s$^{-1}$] & [km s$^{-1}$] & [km s$^{-1}$] & &  & P & R & FS & P & R & FS  & P & R & FS
            \\
            \cmidrule(lr){1-8} \cmidrule(lr){9-11} \cmidrule(lr){12-14} \cmidrule(lr){15-17} 
            D1 & Yes &485 & 1008 & 461 & 997 & 0.62 & 0.77 & 0.99 & 0.61 & 0.75 & 0.81 & 1.00 & 0.89 & 0.88 & 0.85 & 0.84\\
            D2 & Yes &346 & 658 & 367 & 697 & 0.63 & 0.74 & 0.93 & 0.66 & 0.77 & 0.83 & 0.97 & 0.90 &0.87 & 0.86 & 0.85\\
            D3\astfootnote{\label{name}Training set} & Yes &362 & 721 & 363 & 748 & 0.62 & 0.78 & 0.98 & 0.56 & 0.71 & 0.78 & 0.99 & 0.88 & 0.86 & 0.83 & 0.82 \\
            D4 &Yes & 275 & 643 & 351 & 730 & 0.82 & 0.97 & 0.86 & 0.15 & 0.26 & 0.85 & 0.99 & 0.92 & 0.85  & 0.85  & 0.80 \\
            D5 &Yes & 240 & 448 & 270 & 493 & 0.63 & 0.62 & 0.76 & 0.80 & 0.78 & 0.88 & 0.86 & 0.87 & 0.84  & 0.84 & 0.84 \\
            D6 &Yes & 423 & 949 & 436 & 964 & 0.73 & 0.80 &0.94 & 0.70 & 0.80 & 0.90 & 0.98 & 0.94 & 0.91 & 0.91 & 0.91 \\
            D7 &No & 852 & 779 & 482 & 983 & 0.02 & 0.77 & 1.00 & 0.24 & 0.38 & 0.02 & 0.99 & 0.04 & 0.98 & 0.24 & 0.37  \\
            D8 &Yes & 580 & 1209 & 703 & 1281 & 0.94 & 0.83 & 0.26 & 0.78 & 0.39 & 0.98 & 0.87 & 0.92 & 0.94 & 0.86 & 0.89 \\
            D9\footref{name} &Yes & 228 & 428 & 228 & 445 & 0.54 & 0.64 & 0.97 & 0.76 & 0.85 & 0.83 & 0.98 & 0.90 & 0.89 & 0.88 & 0.88\\
            D10\footref{name} &Yes & 484 & 1002 & 477 & 1015 & 0.67 & 0.83 & 0.94 & 0.49 & 0.64 & 0.80 & 0.98 & 0.88 & 0.84 & 0.82 & 0.80\\
            D11 & No & 810 & 734 & 410 & 905 & 0.01 & 0.79 & 1.00  & 0.19 & 0.32 & 0.02 & 0.97 & 0.83 & 0.98 & 0.22 & 0.34 \\
            D12 &Yes & 447 & 1012 & 519 & 1091 & 0.73 & 0.77 & 0.84 & 0.75 & 0.79 & 0.91 & 0.95 & 0.93 & 0.89 & 0.89 & 0.89\\
            D13 & No & 1190 & 1145 & 651 & 1283 & 0.02 &  0.92 & 1.00 & 0.08 & 0.15 & 0.02 & 0.99 & 0.04 & 0.98 & 0.10 & 0.14\\
            D14 &Yes & 575 & 1101 & 559 & 1109 & 0.56 &  0.82 & 0.99 & 0.41 & 0.58 & 0.69 & 0.99 & 0.81 & 0.82 & 0.74 & 0.71\\
            D15 &Yes & 381 & 1044 & 529 & 1176 & 0.84 &  0.79 & 0.61 & 0.78 & 0.69 & 0.96 & 0.90 & 0.93 & 0.90 & 0.89 & 0.89 \\
            D16 &Yes & 848 & 1495 & 878 & 1543 & 0.73 &  0.92 & 0.95 & 0.27 & 0.42 & 0.79 & 0.99 & 0.88 & 0.83 & 0.80 & 0.76\\
            D17 &Yes & 535 & 902 & 562 & 940 & 0.67 &  0.81 & 0.91 & 0.69 & 0.78 & 0.81 & 0.97 & 0.89 & 0.85 & 0.83 & 0.82\\
            D18\footref{name} &Yes & 472 & 837 & 475 & 845 & 0.794 &  0.737 & 0.91 & 0.69 & 0.79 & 0.86 & 0.96 & 0.91 & 0.87 & 0.87 & 0.86 \\
            D19 &Yes & 453 & 967 & 435 & 954 & 0.765 &  0.767 & 0.99 & 0.61 & 0.75 & 0.80 &  1.00 & 0.77 & 0.87 & 0.85 & 0.84\\
            D20 &Yes & 570 & 1071 & 596 & 1113 & 0.779 &  0.791 & 0.99 & 0.54 & 0.70 & 0.80 & 1.00 & 0.89 & 0.87 & 0.84 & 0.82\\
            D21 &Yes & 760 & 1333 & 689 & 1261 & 0.738 & 0.817 & 1.00 & 0.39 & 0.57 & 0.70 & 1.00 & 0.83 & 0.81 & 0.73 & 0.68\\
            D22\footref{name} &Yes & 550 & 1147 & 552 & 1173 & 0.62 & 0.87 & 0.99 & 0.34 & 0.50 & 0.71 & 0.99 & 0.83 & 0.82 & 0.75 & 0.71\\
            D23 &Yes & 630 & 1266 & 647 & 1285 & 0.82 &  0.91 & 0.86 & 0.73 & 0.79 & 0.95 & 0.98 & 0.96 & 0.88 & 0.88  & 0.87\\
            D24 &Yes & 552 & 1038 & 466 & 977 & 0.52 & 0.80 &1.00 & 0.42 & 0.59 & 0.65 & 1.00 & 0.79 & 0.82 & 0.73 & 0.70\\
            D25 &Yes & 510 & 870 & 485 & 833 & 0.63 & 0.73 & 0.96 & 0.71 & 0.81 & 0.85 & 0.99 & 0.91 & 0.89 & 0.88 & 0.88\\
            D26 &Yes & 447 & 986 & 489 & 1032 & 0.74 & 0.76 & 0.82 & 0.76 & 0.79 & 0.92 & 0.94 & 0.93 & 0.89 & 0.89 & 0.89\\
            D27 &Yes & 489 & 937 & 495 & 952 & 0.57 & 0.84 &0.98 & 0.40 & 0.57 & 0.70 & 0.99 & 0.82 & 0.82 & 0.74 & 0.71\\
            D28 &Yes & 182 & 1348 & 958 & 1424 & 0.99 & 0.93 & 0.02 & 0.81 & 0.04 & 1.00 & 0.85 & 0.92 & 0.99 & 0.93 & 0.96 \\
            D29 &Yes & 524 & 1007 & 453 & 919 & 0.47 & 0.78 & 1.00 & 0.41 & 0.58 & 0.60 & 1.00 & 0.77 & 0.81 & 0.69 & 0.66\\
            \\
    
            \caption{Results from the best-fit procedure applied to the double Maxwellian in both the ICL-Subfind and ML algorithm case. We list the convergence report of ICL-Subfind in the second column. Then, we present the ICL and BCG velocity dispersions, the fraction of ICL, and in the last columns the algorithm performance scores (of the two classes and their weighted median), namely: the precision P, recall R, and F-Score FS.}
            \label{tab:compar_ml-subfind}
        \end{longtable}
        
\end{landscape}
\twocolumn

\section*{Acknowledgements}

We would like to thank Massimo Brescia for useful discussions. The simulations presented in this paper have been carried out: at CINECA, with computing time provided through an ISCRA-B project, CINECA-INAF and CINECA-UNITS agreement; at the computing centre of INAF-Osservatorio Astronomico di Trieste, under the coordination of the CHIPP project \citep[][]{bertocco2019inaf,taffoni2020chipp}. This paper has been partially supported by the INDARK INFN Grant. AS and IM are supported by ERC-StG ‘ClustersXCosmo’ grant agreement 716762, by the FARE-MIUR grant ’ClustersXEuclid’ R165SBKTMA.

\section*{Data Availability}

The data underlying this article will be shared on reasonable request to the corresponding author.



\bibliographystyle{mnras}
\bibliography{Bibliography} 



\appendix
\section{Hints for the identification of shell galaxies}
\label{Appendix}

\begin{figure*}
\centering
    \includegraphics[scale=0.45,angle=0.0]{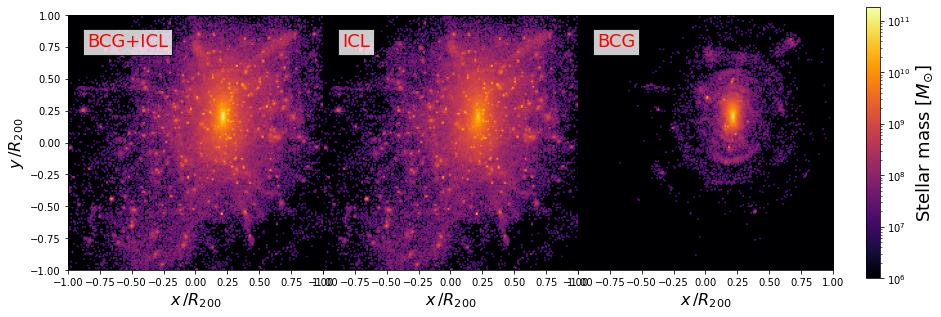}
    \caption{Mass-weighted map of the stellar components in cluster D7: ICL+BCG (in the left panel), ICL (central) and BCG (left) spatial distributions. We only show the results from the predicted labels, since the ICL-Subfind does not converge for this case.}
    \label{fig:D7_maps}
\end{figure*}

Among the test cluster set, we observe a handful of clusters that present particularly interesting features in their spatial distribution. One of the most striking cases is reported on the mass-weighted map in Fig. \ref{fig:D7_maps}. We point out that this cluster has not reached convergence in the ICL-Subfind algorithm, therefore our next discussion will be mostly addressed to the ML output. In the central panel, we display both components (BCG+ICL), the central one illustrates the ICL population, while the right plot is for the BCG stars. The plot shows a complex stellar structure composed of spherical shells surrounding the central stellar peak in the BCG, which is not seen in the ICL-Subfind case. We assume this feature to not be caused by some numerical artefact, given that several clusters have this same symmetrical distribution in the ICL-Subfind analysis. This shell-like distribution could be due to the expansion and later disruption of stars occurring during a tidal shock, which gives origin to shell galaxies. The shells are formed as density waves induced in a thick disc population of dynamically cold stars by a weak interaction with another galaxy \cite{thomson1991shell} and relatively major mergers \citep [e.g., with a mass ratio of 1:10][]{pop2018formation}. It is not clear whether the complexity of the shell structure may be responsible for the difficulty of ICL-Subfind converging, but it is striking to notice that the ML algorithm can detect it without specific training on this particular feature. Additionally, it is also able to spot it whenever the ICL-Subfind does. We will defer a complete analysis of this hypothesis to future work. 

\bsp	
\label{lastpage}
\end{document}